\documentclass{emulateapj}
\usepackage{amsmath}
\usepackage{epstopdf}

\newcommand{\kmps}{\ensuremath{\mathrm{km}\ \mathrm{s}^{-1}}}
\newcommand{\pcc}{\ensuremath{\mathrm{cm}^{-3}}}
\newcommand{\Msol}{\ensuremath{M_{\odot}}}
\newcommand{\Zsol}{\ensuremath{Z_{\odot}}}
\newcommand{\Zobs}{\ensuremath{Z_\mathrm{obs}}}
\newcommand{\nhalo}{\ensuremath{n_\mathrm{H,h}}}
\newcommand{\Thalo}{\ensuremath{T_\mathrm{h}}}
\newcommand{\Zhalo}{\ensuremath{Z_\mathrm{h}}}
\newcommand{\fhalo}{\ensuremath{f_\mathrm{h}}}
\newcommand{\Fhalo}{\ensuremath{F_\mathrm{h}}}
\newcommand{\ncl}{\ensuremath{n_\mathrm{H,cl}}}
\newcommand{\Tcl}{\ensuremath{T_\mathrm{cl}}}
\newcommand{\NHHVC}{\ensuremath{N_\mathrm{H,HVC}}}
\newcommand{\rcl}{\ensuremath{r_\mathrm{cl}}}
\newcommand{\vcl}{\ensuremath{\boldsymbol{v}_\mathrm{cl}}}
\newcommand{\Zcl}{\ensuremath{Z_\mathrm{cl}}}
\newcommand{\Zcool}{\ensuremath{Z_\mathrm{cool}}}

\shorttitle{}
\shortauthors{HENLEY ET AL.}

\begin{document}

\title{The Effect of Mixing on the Observed Metallicity of the Smith Cloud}
\author{David B. Henley, Jeffrey A. Gritton, and Robin L. Shelton}
\affil{Department of Physics and Astronomy, University of Georgia, Athens, GA 30602 \\
  dbh@physast.uga.edu, jgritton@uga.edu, rls@physast.uga.edu}

\begin{abstract}
Measurements of high-velocity clouds' metallicities provide important clues about their origins, and
hence on whether they play a role in fueling ongoing star formation in the Galaxy.  However,
accurate interpretation of these measurements requires compensating for the galactic material that
has been mixed into the clouds.  In order to determine how much the metallicity changes as a result
of this mixing, we have carried out three-dimensional wind-tunnel-like hydrodynamical simulations of
an example cloud.  Our model cloud is patterned after the Smith Cloud, a particularly well-studied
cloud of mass $\sim5\times10^6~\Msol$.  We calculated the fraction of the high-velocity material
that had originated in the galactic halo, \Fhalo, for various sight lines passing through our model
cloud.  We find that \Fhalo\ generally increases with distance from the head of the cloud, reaching
$\sim$0.5 in the tail of the cloud.  Models in which the metallicities (relative to solar) of the
original cloud, \Zcl, and of the halo, \Zhalo, are in the approximate ranges $0.1\la\Zcl\la0.3$ and
$0.7\la\Zhalo\la1.0$, respectively, are in rough agreement with the observations. Models with
$\Zhalo\sim0.1$ and $\Zcl\ga0.5$ are also in rough agreement with the observations, but such a low
halo metallicity is inconsistent with recent independent measurements.  We conclude that the Smith
Cloud's observed metallicity may not be a true reflection of its original metallicity and that the
cloud's ultimate origin remains uncertain.
\end{abstract}


\section{INTRODUCTION}
\label{sec:Introduction}

High-velocity clouds (HVCs) are clouds in the Galactic halo traveling at high speed relative to the
local standard of rest ($\ga$90~\kmps; \citealt{wakker97}). HVCs play an important role in the
Galaxy, by supplying material to fuel star formation. However, their origins are uncertain, with
Galactic fountains, gas stripped from satellite galaxies, and remnants from the formation of the
Local Group all having been suggested as possible origins (see reviews by \citealt{bregman04} and
\citealt{putman12}). The metallicity of an HVC is key to determining its origin, as it allows us to
distinguish between material that has or has not been processed through the Galactic disk.

The Smith Cloud \citep[SC;][]{smith63} is arguably the best-characterized of the Galaxy's HVCs.  Its
distance and mass are both known: $12.4 \pm 1.3$~kpc \citep[hereafter L08]{lockman08} and
$\ga10^6~\Msol$ each in \ion{H}{1} and \ion{H}{2} (L08; \citealt[hereafter H09]{hill09};
\citealt{lockman15}), respectively. The cloud is located approximately 3~kpc from the Galactic
midplane, is approximately 1~kpc $\times$ 3~kpc on the sky, and is thought to be moving in the
direction of its major axis. Uniquely among HVCs, its three-dimensional velocity is also
constrained. From the variation in the line-of-sight velocity across the cloud, L08 concluded that
the cloud is approaching the disk at $\sim$70~\kmps\ on a prograde orbit $\sim$50~\kmps\ faster than
the Galactic rotation. Based purely on its orbit, L08 were unable to favor a Galactic or
extragalactic origin for the SC, and pointed out the need to measure its metallicity.

Using measurements of the [\ion{N}{2}] emission from the SC's head, H09 determined that the cloud's
nitrogen abundance is likely 0.15--0.44 times solar, and concluded that this low abundance supported the
idea that the cloud represents new material that is being accreted onto the Galaxy. On the other
hand, using \ion{S}{2} absorption lines observed along three sight lines passing through the cloud's
tail, \citet[hereafter F16]{fox16} measured the cloud's sulfur abundance to be $0.53^{+0.21}_{-0.15}$
times solar. Based on this, and the orbit determined by L08, F16 concluded that the SC had a
Galactic origin (specifically, in the outer disk $\sim$13~kpc from Galactic Center, where the
metallicity is half solar).

It is important to note that the cloud should have experienced some mixing with the surrounding gas.
Using hydrodynamical simulations, \citet{gritton14} showed that the high-velocity material in the
tail of an HVC may contain significant amounts of material that originated in the halo as a result.
Therefore, the true metallicity of the original cloud is not simply given by measurements of the
metallicity of the high-velocity material, but depends also on the halo metallicity and the degree
of mixing between the halo and the cloud.

The goal of the current study is to determine how extensively mixing has changed the metallicity of
an example HVC, the SC. To this end, we present three-dimensional hydrodynamical simulations of an
SC-like HVC traveling through the halo (Section~\ref{sec:HydroSims}).  We use these simulations to
quantify the degree of mixing of cloud and halo material (Section~\ref{sec:Mixing}), and examine the
relationship between the resulting metallicites and H09's and F16's metallicity measurements
(Section~\ref{sec:Comparison}). We discuss our results in Section~\ref{sec:Discussion}.

\begin{deluxetable*}{lcc}
\tablecaption{Reference Model (Model A) Parameters\label{tab:ModelParameters}}
\tablehead{
\colhead{Quantity}            & \colhead{Symbol}      & \colhead{Value}
}
\startdata
Domain $x$ range              &                       & $-2.56$ to +2.56~kpc  \\
Domain $y$ range              &                       & 0 to +2.56~kpc        \\
Domain $z$ range              &                       & $-1.28$ to +14.08~kpc \\
Spatial resolution\tablenotemark{a,b}
                              &                       & 10 to 160~pc          \\
Cooling curve metallicity\tablenotemark{c}
                              & \Zcool                & $10^{-0.5} \Zsol$       \\
Cloud radius                  & \rcl                  & 500~pc                \\
Cloud hydrogen density        & \ncl                  & 0.4~\pcc              \\
Cloud temperature\tablenotemark{d}
                              & \Tcl                  & 1500~K                \\
Cloud speed                   & $|\vcl|$              & 150~\kmps             \\
Ambient halo temperature      & \Thalo                & $2\times10^{6}$~K      \\
Ambient halo hydrogen density\tablenotemark{e}
                              & \nhalo                & $3\times10^{-4}$~\pcc  \\
\enddata
\tablenotetext{a}{We used adaptive mesh refinement during the simulation. If the whole domain were
  modeled at full resolution, the grid would consist of $512\times256\times1536$
  cells, each of size (10~pc)$^3$.}
\tablenotetext{b}{In models Alr and Ahr, the maximum resolution was 20 and 5~pc, respectively.}
\tablenotetext{c}{In Model~B, we used $\Zcool=\Zsol$.}
\tablenotetext{d}{Set by requiring pressure balance between the cloud and the ambient medium.
    In Model~C, $\Tcl=2500$~K.}
\tablenotetext{e}{In Model~C, $\nhalo=5\times10^{-4}~\pcc$.}
\end{deluxetable*}

\section{HYDRODYNAMICAL SIMULATIONS}
\label{sec:HydroSims}

\subsection{Model Description}

We used FLASH \citep{fryxell00} version 4.2 to simulate the hydrodynamical interaction between our
model cloud and the halo. We used a three-dimensional Cartesian domain. Our simulations are similar
to Model~B in \citet{gritton14}, in that the cloud started at rest at the origin, and the ambient
medium flowed upward through the domain with velocity $-\vcl=|\vcl|\boldsymbol{\hat{z}}$, where
\vcl\ is the cloud's velocity in the ISM rest frame. This enabled us to study the cloud's evolution
over a long period of time, without the need for a prohibitively large domain. However, whereas
\citet{gritton14} modeled a quarter-cloud, assuming symmetry about the $x=0$ and $y=0$ planes, here
we modeled a half-cloud, assuming symmetry only about the $y=0$ plane. The size of our model domain
is summarized in Table~\ref{tab:ModelParameters}, as are the other model parameters and notation,
discussed below.

Our reference model is called Model~A. Following \citet{galyardt16}, we initialized the cloud as a
sphere of radius $\rcl=500$~pc with hydrogen number density $\ncl=0.4~\pcc$ and hydrogen to helium
ratio of 10 to 1. Note that, at maximum resolution, the cloud's diameter is covered by 100 cells
(for an example comparison, the corresponding number of cells from \citealt{gritton14} is 33).  The
cloud's density profile follows a hyperbolic tangent function at the cloud edge
\citep{kwak11,gritton14}.  This yields a total hydrogen mass of $5\times10^6~\Msol$, similar to the
observed mass of the SC (L08; H09). Also, the model cloud's initial diameter is equal to the
observed length of the SC's minor axis ($\sim$1~kpc; note that the cloud is thought to be moving in
the direction of its major axis; L08).  The model cloud's initial speed relative to the ISM was
150~\kmps\ (i.e., the ambient medium flowed in the $+z$ direction at 150~\kmps). Although this is
slightly higher than the SC's current speed relative to the ISM ($130\pm14$~\kmps; L08), the ram
pressure of the ISM causes the cloud to decelerate relative to the ISM during the course of the
simulation.

We assumed that the temperature of the ambient halo is $\Thalo=2\times10^6$~K
\citep[e.g.,][]{henley13,henley15c}. The density of the halo is uncertain, though various
observational constraints imply that it does not exceed a few times $10^{-4}~\pcc$, except in the
lower halo (see, e.g., discussion in \citealt{kwak11}, Section~2 and \citealt{henley14b},
Section~4.1).  \citet{miller13} used \ion{O}{7} absorption line measurements to model the halo
density. Their best-fit model implies that the halo density at the SC's current Galactocentric
distance (L08) is $6\times10^{-4}~\pcc$.  If we follow L08 and use the cloud's current position and
velocity to calculate its past trajectory within the Galactic potential of \citet{wolfire95b}
(neglecting drag), we find that the cloud would have experienced an ambient density of
$\sim$$(\mbox{3--6})\times10^{-4}~\pcc$ over the past $\sim$60~Myr (cf., on this trajectory, the
cloud would have been in the disk $\sim$70~Myr ago; L08).  For our reference model, we assumed a
constant ambient density of $\nhalo=3\times10^{-4}~\pcc$.  The temperature of the cloud (1500~K in
Model~A) was set by requiring pressure balance between the cloud and the ambient medium.

Radiative cooling depends on the metallicity of the gas, which is uncertain in the SC and in its
environment. As the halo's metallicity is likely to be subsolar, for our reference model we used the
\citet{sutherland93} cooling curve with metallicity $\log(\Zcool/\Zsol) = -0.5$ (relative to the
\citealt{anders89} solar abundances). Note that radiative cooling was suppressed below $10^4$~K
(i.e., the cooling curve was set to zero below this temperature). This prevents runaway cooling of
the coolest, densest gas. The physical justification for such a cut-off is that the cool gas will be
heated by the ambient ultraviolet radiation field, by photoemission from dust grains
\citep{wolfire95a}. However, we did not attempt to model this heating, and we acknowledge that this
heating rate will be lower above the disk (where our model cloud resides) than in the disk
\citep{wolfire95b}. The simulation time step was limited by multiple criteria and at no time was
greater than half of the cooling time in any cell in the model domain.

In order to test the sensitivity of our results to the assumed cooling curve, we tried a model that
used a solar-metallicity cooling curve (Model~B).  To test the sensitivity to the ambient gas
density, we tried a variant with $\log(\Zcool/\Zsol) = -0.5$ but a higher ambient density,
$\nhalo=5\times10^{-4}~\pcc$ (Model~C).\footnote{We also experimented with a model with
  $\nhalo=6\times10^{-4}~\pcc$, which is equal to the halo density at the SC's current location
  (using the \citealt{miller13} halo model). However, we found that the resulting higher ram
  pressure of the ambient medium braked the high-velocity material too severely, meaning that there
  was virtually no high-velocity material at angles $\ga$25\degr\ behind the cloud head (cf.\ F16's
  sight lines lie 14--28\degr\ behind the SC head).\label{fn:HighDensity}} Finally, to test the
sensitivity to the simulation resolution, we tried two variants in which the maximum resolution was
twice or half that of our reference model (Models~Ahr (high resolution) and Alr (low resolution),
respectively).

We traced the mixing of cloud and halo material by defining two inert fluids which were advected
with the flow, one representing the cloud material and the other the ambient material. The mass
fractions of these fluids were initially one and zero, respectively, within the cloud, and vice
versa outside the cloud. When postprocessing the simulations, we determined the fraction of the
material in each grid cell that originated in the halo, \fhalo, directly from these fluids' mass
fractions.

We ran the simulations for 400~Myr. This is several times longer than the time the SC is expected to
have spent in the halo, assuming it has stayed on its current trajectory (L08). However, running our
simulation for this length of time ensures that the mixing in the model cloud's tail has reached a
reasonably steady state (see Section~\ref{sec:Mixing}).  It also allows us to quantify the time
variability of that mixing, which we use to estimate the uncertainty in the degree of mixing at the
current epoch.

Our simulations do not include a magnetic field or the possibility that the SC may have a dark
matter minihalo, both of which would affect the degree of mixing in the tail.  It is not clear how
significant these omissions are.  Nor do our simulations collide the SC with the Galactic disk
\citep{nichols14}, a scenario that \citet{galyardt16} found would destroy the cloud even if the SC
has a dark matter minihalo.

\subsection{Results}

\begin{figure*}
\centering
\plotone{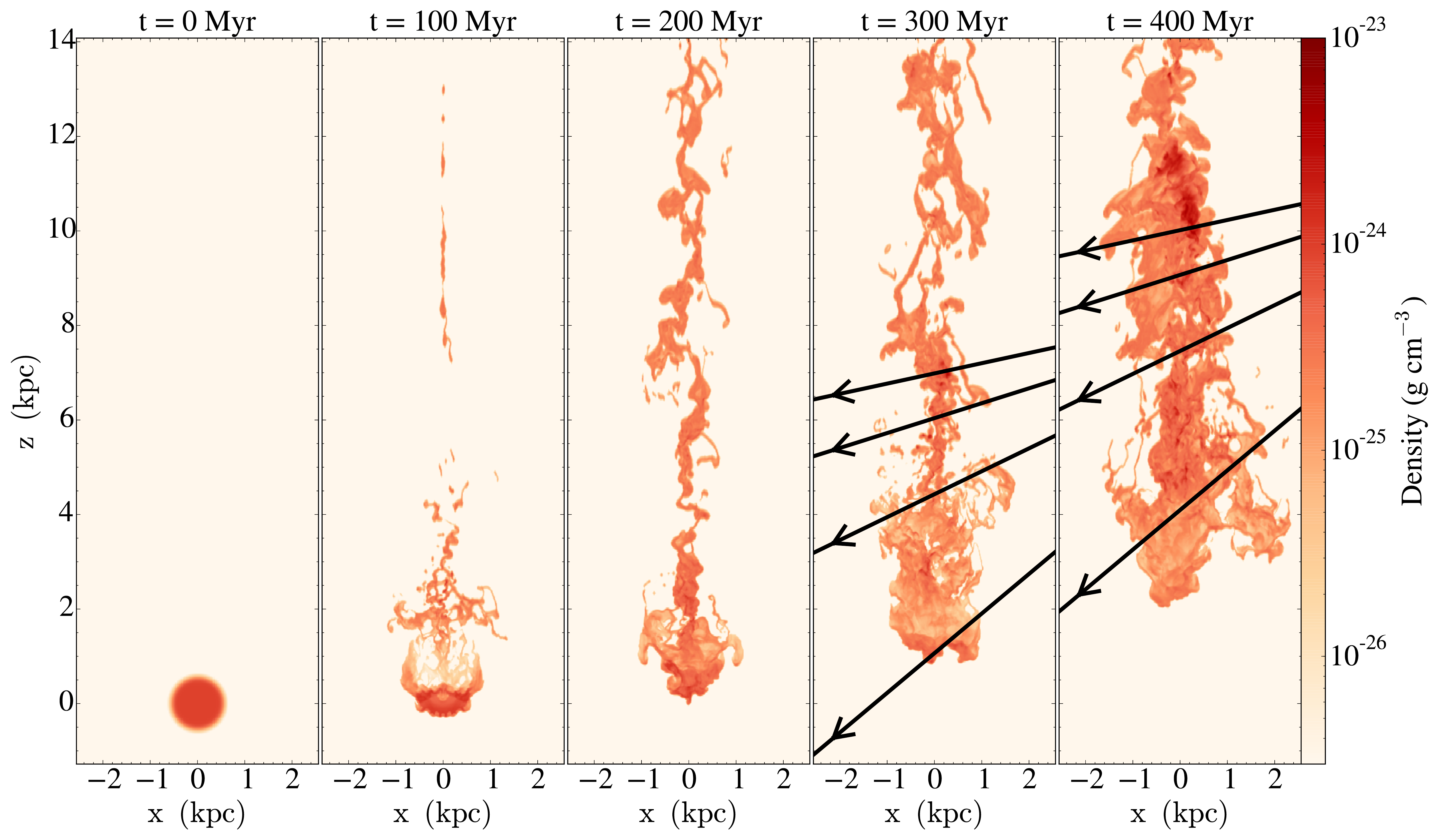}
\caption{Slices through our reference model domain, showing the density in the $xz$ plane at $t=0$,
  100, ..., 400~Myr (left to right). The diagonal black lines in the later epochs' plots show
  model sight lines used to calculate the fraction of the high-velocity material that
  originated in the halo (Section~\ref{sec:Mixing}). These sight lines are from a vantage
  point to the right of the model domain, 12.4~kpc from the cloud head, and cross the $z$ axis
  at angles corresponding to the observational sight lines in Table~\ref{tab:Obs}.
  \label{fig:SliceA}}
\end{figure*}

\begin{figure*}
\centering
\plotone{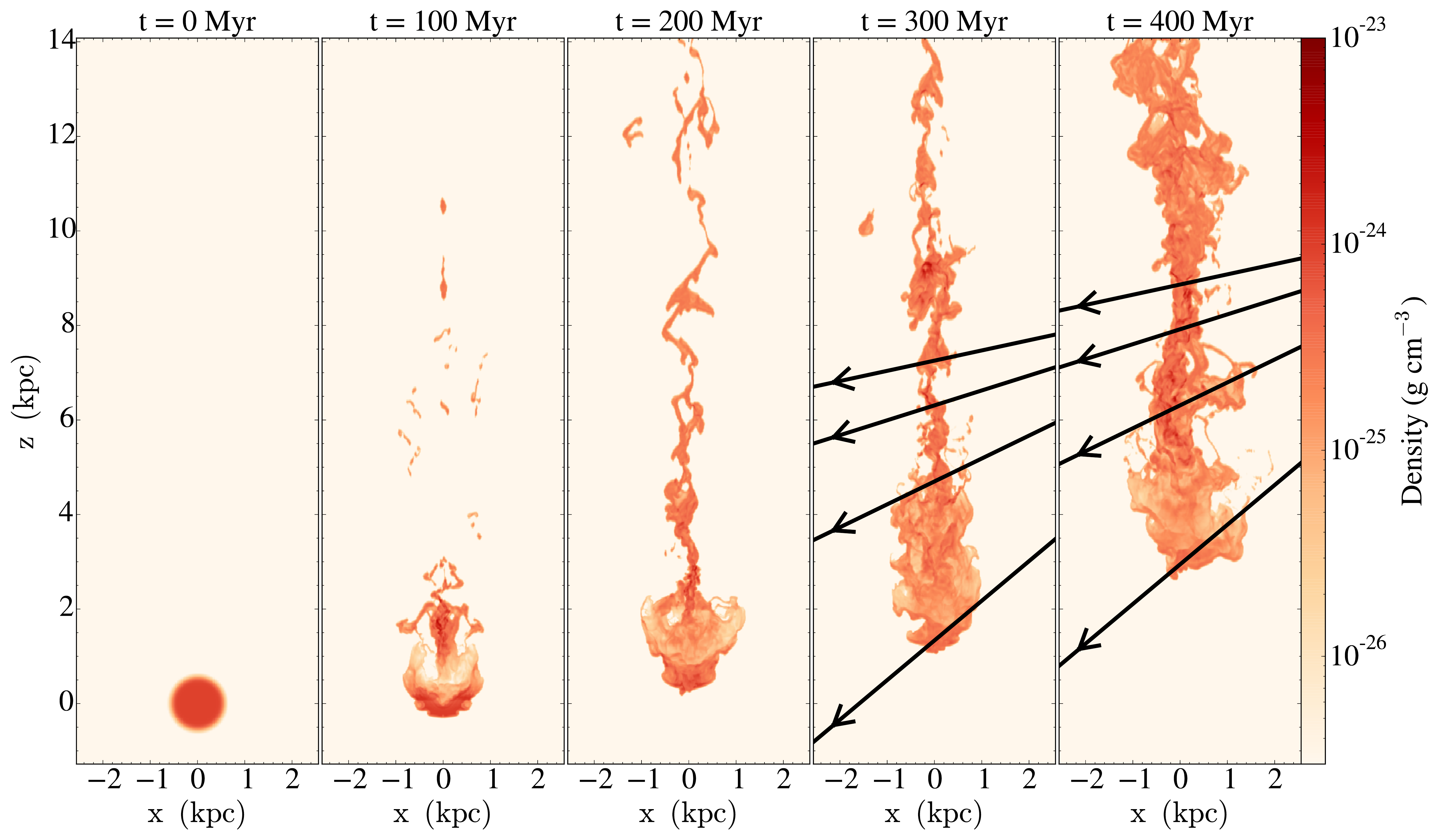}
\caption{Same as Figure~\ref{fig:SliceA}, but for Model~B.
  \label{fig:SliceB}}
\end{figure*}

\begin{figure*}
\centering
\plotone{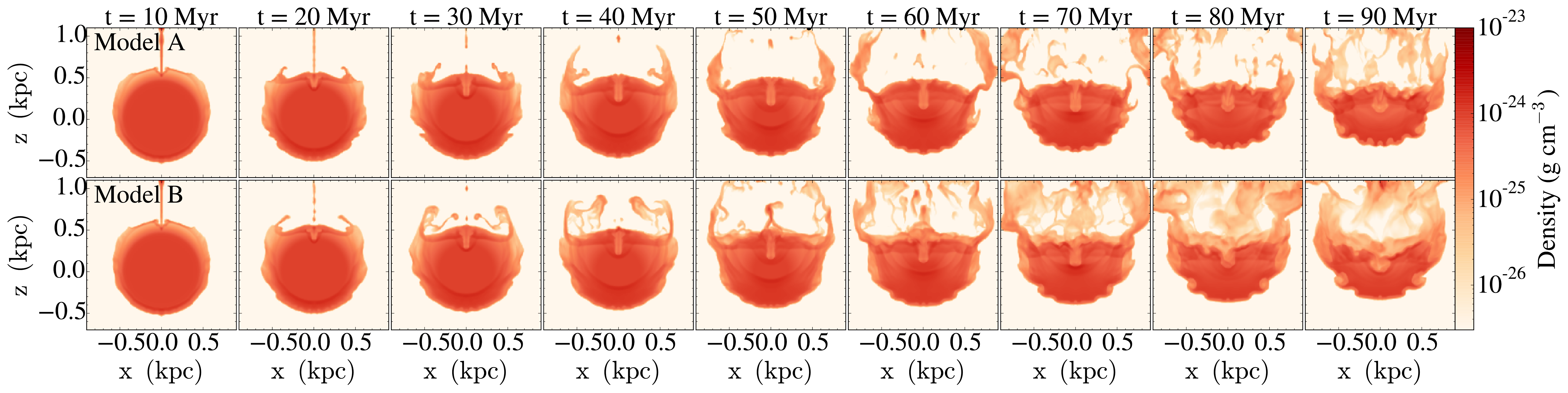}
\caption{Comparison of the densities in the $xz$ plane near the origin from Models~A (upper row) and
  B (lower row), at $t=10$, 20, ..., 90~Myr (left to right).
   \label{fig:SliceAB}}
\end{figure*}

\begin{figure*}
\centering \plotone{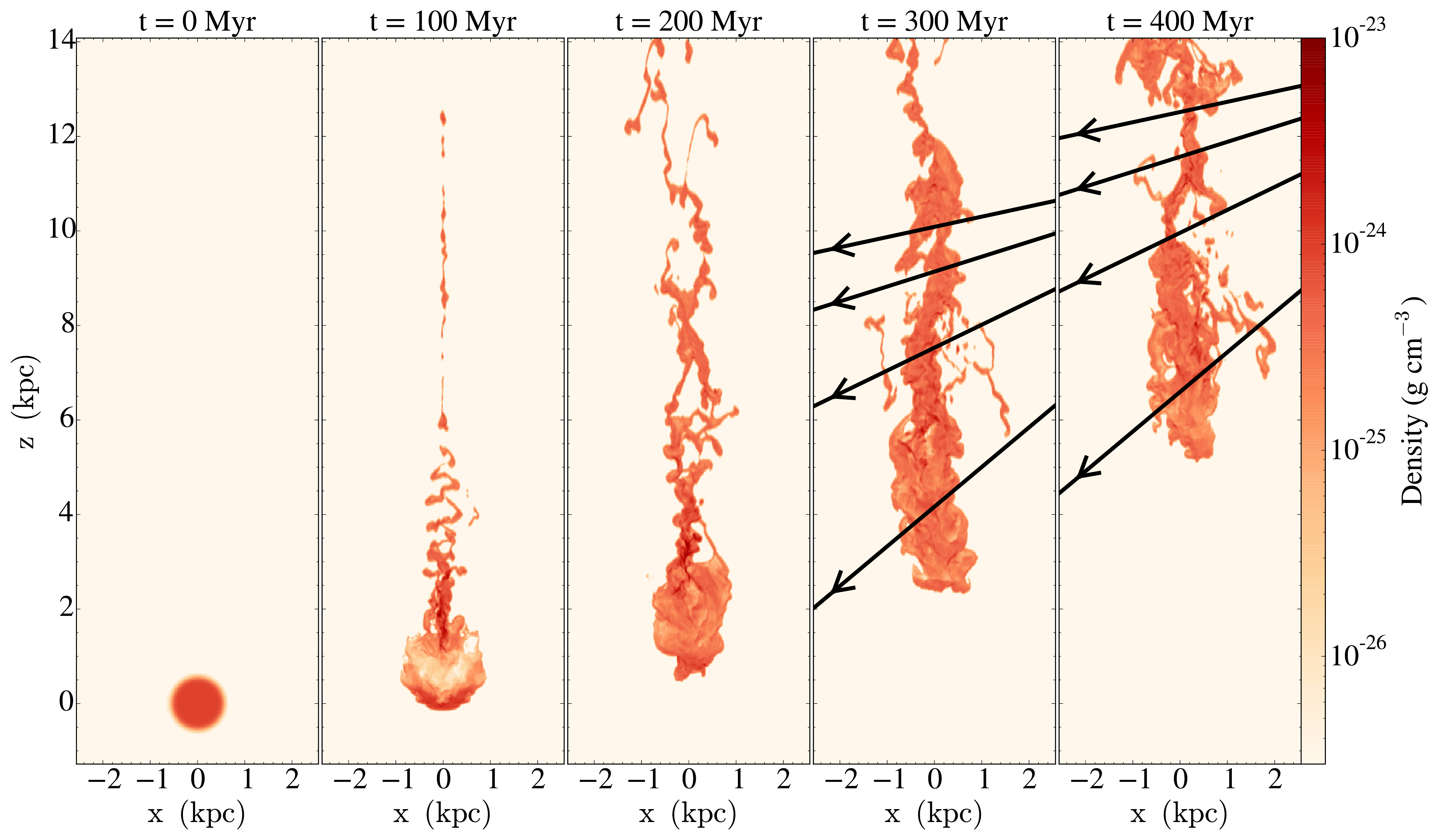}
\caption{Same as Figure~\ref{fig:SliceA}, but for Model~C.
  \label{fig:SliceC}}
\end{figure*}

Figure~\ref{fig:SliceA} shows slices through the reference model domain, showing the density at
100-Myr intervals. The upward-flowing ambient medium ablates material from the cloud, which mixes
with the halo gas and gets stretched out in the $+z$ direction. There is considerable spatial and
temporal variation in the structure of the tail.

When we compare Model~B (Figure~\ref{fig:SliceB}) with Model~A, we find that the Model~A cloud
fragments more than the Model~B cloud. This difference appears to originate in the first
$\sim$90~Myr of the clouds' evolution.  At $t\sim20$~Myr, a vortex emerges from the back of each
cloud, causing the hot ambient medium to be mixed in behind the cloud (Figure~\ref{fig:SliceAB}),
forming intermediate-temperature gas. In Model~B, starting from $t\sim60$~Myr, this gas cools and
becomes denser than in Model~A, because of Model~B's larger cooling rate.  This tends to smooth out
the density structure behind the cloud, which impairs the fragmentation of the cloud, compared with
Model~A.

In Model~C (Figure~\ref{fig:SliceC}), the cloud is pushed further up through the model domain than
in Model~A, because of the greater ram pressure of the ambient medium.  (At $t=400$~Myr, the cloud
head is at $z=6.6$ and 4.1~kpc in Models~C and A, respectively.) The Model~C cloud also fragments
less than the Model~A cloud. This result is surprising, as the timescale for the growth of
Kelvin-Helmholtz instabilities is expected to be smaller in Model~C, due to the smaller density
contrast between the cloud and the ambient gas \citep{chandrasekhar61}.

\section{MIXING OF HALO AND CLOUD MATERIAL}
\label{sec:Mixing}

\begin{deluxetable*}{lcccccccc}
\tablecaption{Observation Details and Model Results\label{tab:Obs}}
\tablehead{
\colhead{Target}        & \colhead{$l$}   & \colhead{$b$}   & \colhead{Dist. from SC head\tablenotemark{a}} & \colhead{$\log(Z/\Zsol)$\tablenotemark{b}} & \colhead{Ref.} & \multicolumn{3}{c}{\Fhalo\tablenotemark{c}} \\
\cline{7-9}
                        & \colhead{(deg)} & \colhead{(deg)} & \colhead{(deg)}                               &                                            &                & \colhead{Model A} & \colhead{Model B} & \colhead{Model C}
}
\startdata
Cloud head              & 38.6            & $-13.1$         & 0\tablenotemark{d}                            & $-0.82$ to $-0.36$                         & H09            & $0.153 \pm 0.030$ & $0.132 \pm 0.017$ & $0.217 \pm 0.038$ \\
RX J2043.1+0324         & 49.72           & $-22.88$        & 14.13                                         & $-0.14 \pm 0.20$                           & F16            & $0.449 \pm 0.047$ & $0.292 \pm 0.043$ & $0.414 \pm 0.048$ \\
PG 2112+059             & 57.04           & $-28.01$        & 22.49                                         & $-0.09 \pm 0.36$                           & F16            & $0.493 \pm 0.058$ & $0.426 \pm 0.066$ & $0.555 \pm 0.144$ \\
RX J2139.7+0246         & 58.09           & $-35.01$        & 27.82                                         & $-0.58 \pm 0.25$                           & F16            & $0.535 \pm 0.076$ & $0.503 \pm 0.085$ & $0.692 \pm 0.166$ \\
\enddata
\tablenotetext{a}{Located at $(l,b)=(38\fdg67,-13\fdg41)$ (L08).}
\tablenotetext{b}{SC metallicity, assuming that H09's nitrogen abundance and F16's sulfur abundances
  correspond directly to the overall metallicity, i.e.,
  $[\mathrm{N}/\mathrm{H}]=[\mathrm{S}/\mathrm{H}]=\log(Z/\Zsol)$.  For the F16 measurements we have
  combined the statistical and systematic uncertainties.}
\tablenotetext{c}{Fraction of high-velocity material along the sight line originating in the halo,
  extracted from our hydrodynamical models.}
\tablenotetext{d}{The tabulated coordinates give the pointing direction of the Wisconsin H$\alpha$ Mapper (WHAM), used by H09.
  Although this direction is not exactly toward the location of the SC head given in L08, this location lies within the
  WHAM field of view.}
\end{deluxetable*}

\begin{figure}
\centering
\plotone{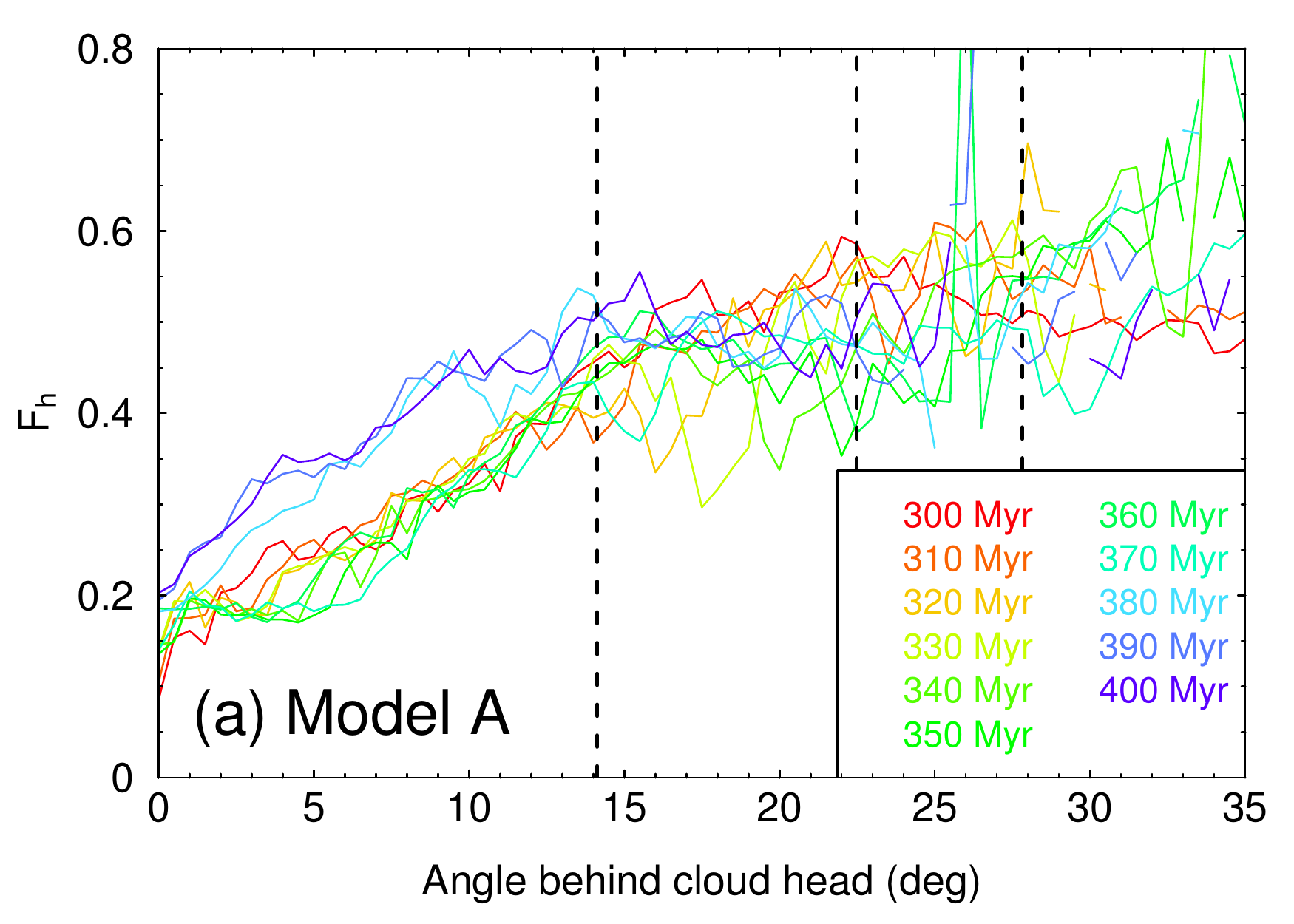}
\plotone{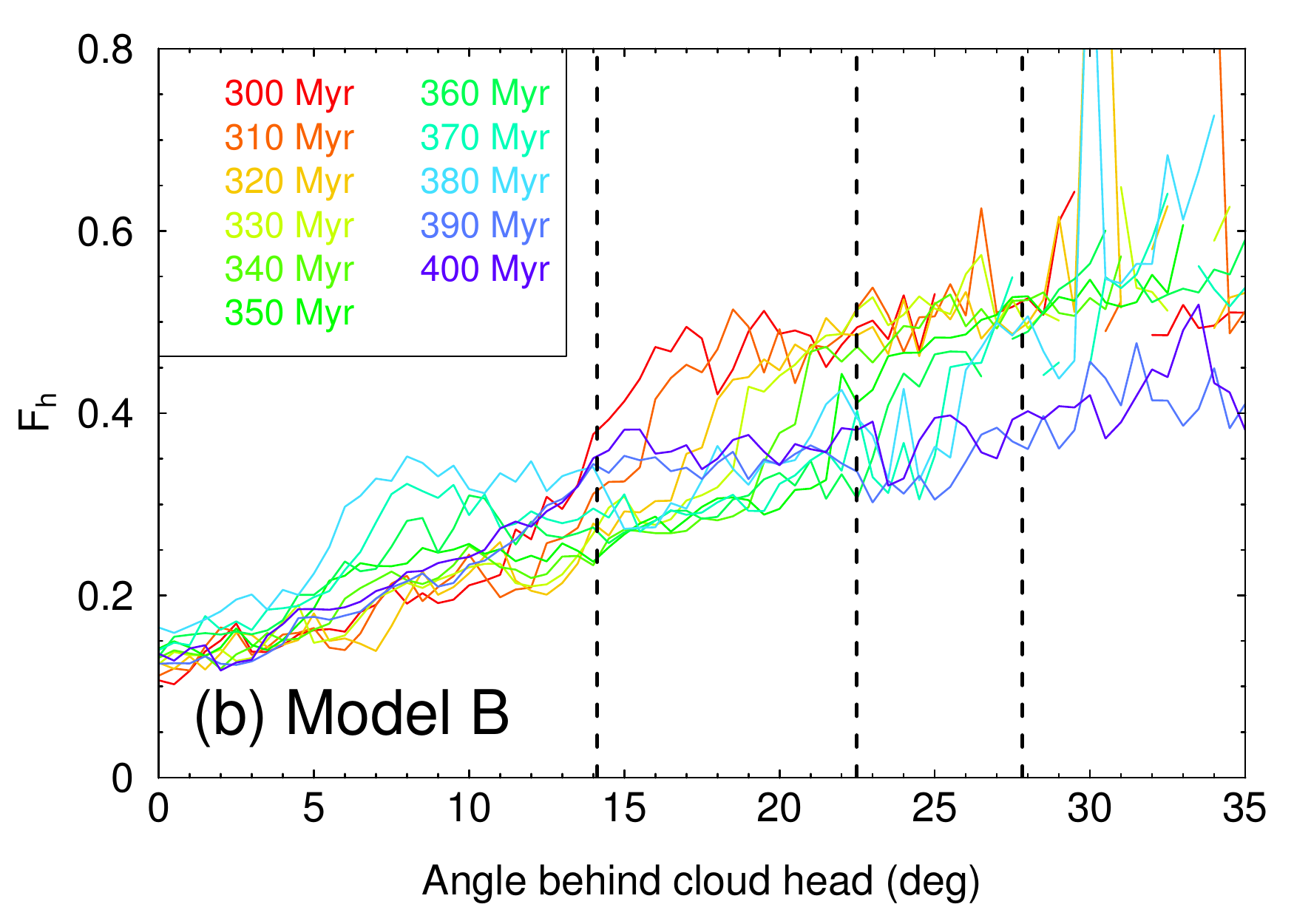}
\plotone{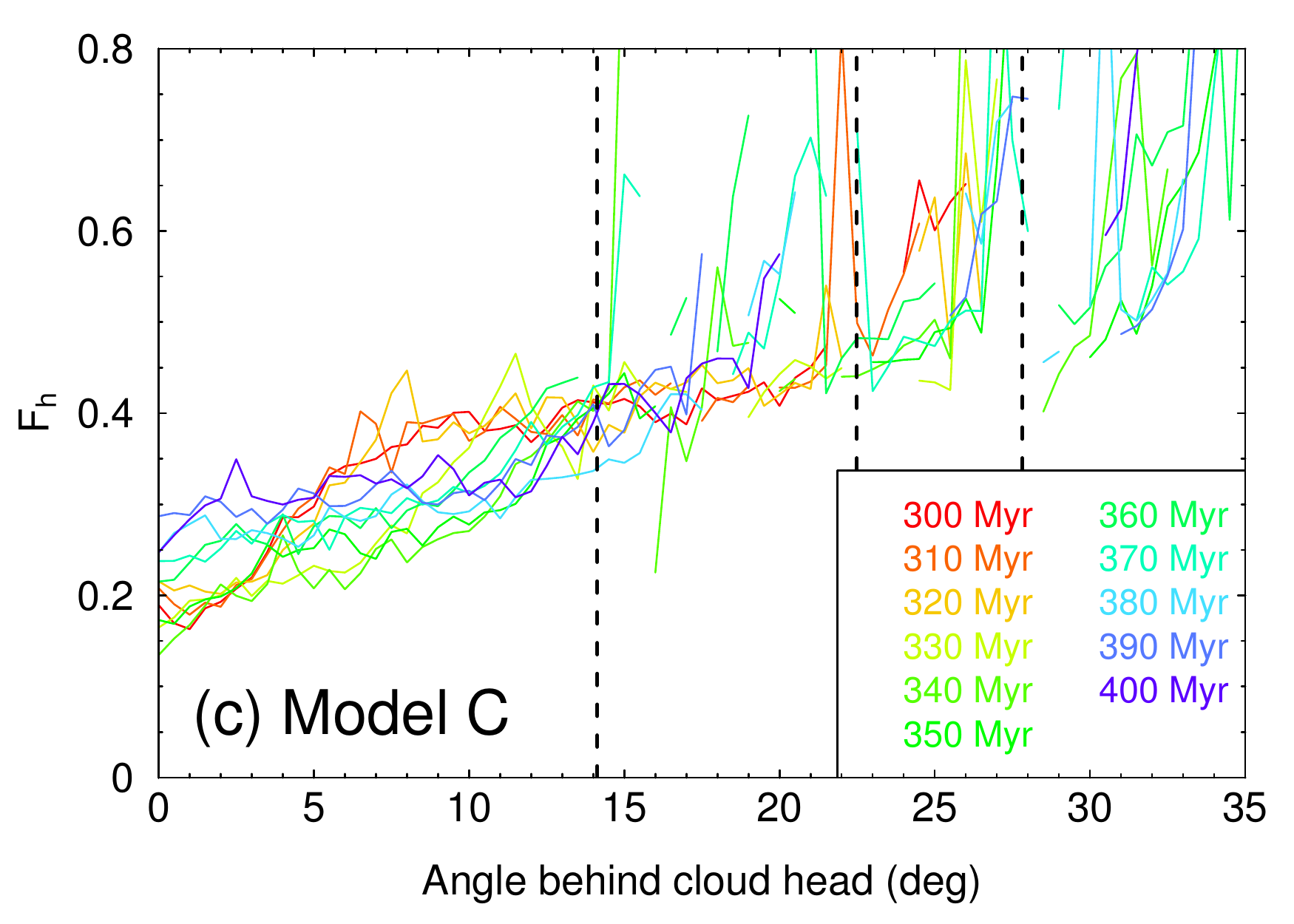}
\caption{The fraction of the high-velocity material originating in the halo, \Fhalo, as a function
  of angular distance behind the cloud head, extracted from various epochs of Models~A--C (top to
  bottom; see keys).
  The vertical dashed lines indicate the location of F16's sight lines relative to the SC head.
  \label{fig:MixingProfile}}
\end{figure}

We calculated the proportion of halo material within the high-velocity gas on various sight lines
running through our model domain in the following manner.  We first defined high-velocity material
as that which travels faster than 90~\kmps\ relative to the ambient medium.  We then defined
\fhalo\ for any given cell in the simulational domain as the fraction of the high-velocity material
in that cell that originated in the halo.  Thus, $1-\fhalo$ is the fraction of the high-velocity
material in the cell that originated in the HVC.  For any sight line through the simulational
domain, the fraction of the high-velocity material that originated in the halo is
\begin{equation}
  \Fhalo = \frac{\int \fhalo \rho \, dl}{\int \rho \, dl},
  \label{eq:Fhalo}
\end{equation}
where $\rho$ is the gas density and only high-velocity gas is included in the integrals.

We then determined the angle that the simulational domain must be viewed at so as to mimic actual
observations of the SC's head and tail.  This angle is equal to the angle between the SC's velocity
relative to the surrounding co-rotating ISM and the line of sight to the SC's head, which, based on
the cloud's position, distance, and velocity from L08, is 130\degr.

We determined where the cloud's head is located, when viewed from this angle, by choosing the line
of sight that has the greatest high-velocity hydrogen column density, \NHHVC, from all of the lines
of sight that cross the $z$ axis of our domain at 130\degr.  This was done at every model epoch in
order to determine the location of the cloud's head as a function of time.\footnote{At some epochs,
  dense regions would form temporarily in the tail of the model HVC, meaning that the maximum of
  \NHHVC\ was several kiloparsecs behind the front of the model HVC. Therefore, in order to prevent
  one of these temporary dense regions in the model cloud's tail being defined as the cloud head, we
  restricted our search for the maximum of \NHHVC\ to within 2~kpc of the cloud front.}

We examined a series of sight lines that cross through the $z$ axis of the domain behind the cloud's
head.  The angles between each of these sight lines and the sight line through the cloud's head were
determined from the assumption that the observer was on Earth, 12.4~kpc (L08) from the SC (see
Figures~\ref{fig:SliceA}, \ref{fig:SliceB}, and \ref{fig:SliceC}). Figure~\ref{fig:MixingProfile}(a)
shows \Fhalo\ as a function of this angle for Model~A at various epochs.  For angles $\la$30\degr,
\Fhalo\ reaches a reasonably steady state by $t=300$~Myr (i.e., the variation of \Fhalo\ from epoch
to epoch is relatively small compared to earlier epochs). However, there is still some time
variation of the \Fhalo\ profiles due to the fact that the flow in the model HVC's tail is not
steady, and for the last few tens of megayears of the simulation, \Fhalo\ is drifting toward larger
values for angles $\la$15\degr. For Models~B and C (Figures~\ref{fig:MixingProfile}(b) and (c),
respectively), we find that \Fhalo\ also reaches a reasonably steady state by around $t=300$~Myr,
although again there is some time variation of the \Fhalo\ profiles. In addition, the Model~C
profiles exhibit gaps at some angles, meaning that there is no material above the high-velocity
cut-off along the sight line at those particular angles. These gaps in the high-velocity material
distribution are due to the higher-density ambient medium having a stronger braking effect on the
cloud material than in the other models (this braking effect was noted in
footnote~\ref{fn:HighDensity} for a model with an even higher ambient density, for which the effect
was more severe than in Model~C).

\begin{figure}
\centering
\plotone{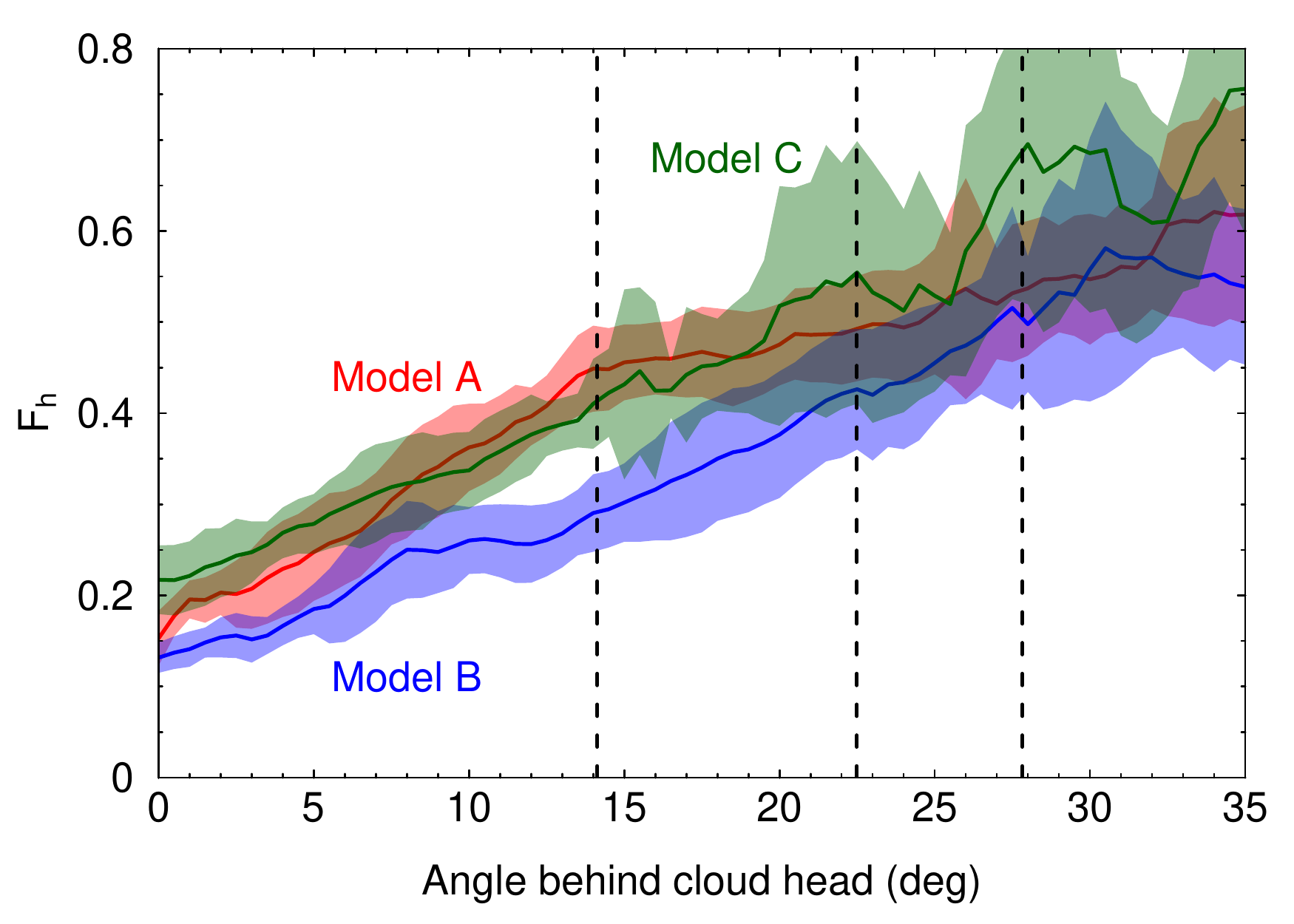}
\caption{\Fhalo\ sampled every 1~Myr and averaged over $t=300$--400~Myr from Models~A, B, and
    C (red, blue, and green, respectively). The colored bands indicate the standard deviation.
  The vertical dashed lines indicate the location of F16's sight lines relative to the SC head.
  \label{fig:AverageProfile}}
\end{figure}

Figure~\ref{fig:AverageProfile} shows \Fhalo\ averaged over multiple epochs. Within the
epoch-to-epoch variations, these averaged profiles are similar for Models~A and B at the cloud head
and at angles $\ga$20\degr\ behind the cloud head. The difference between the models at intermediate
angles is because the Model~A cloud fragments more than the Model~B cloud (as noted above), leading
to more rapid mixing in the first few kiloparsecs behind the cloud head. However, other than in the
first few kiloparsecs behind the cloud head, the mixing results are not very sensitive to the
assumed cooling rate. In contrast, the averaged profiles from Models~A and C are similar at all
angles, though this is in part due to the large time variation of the Model~C profiles. These
results suggest that the extent of mixing is not very sensitive to the assumed ambient density.

\begin{figure}
\centering
\plotone{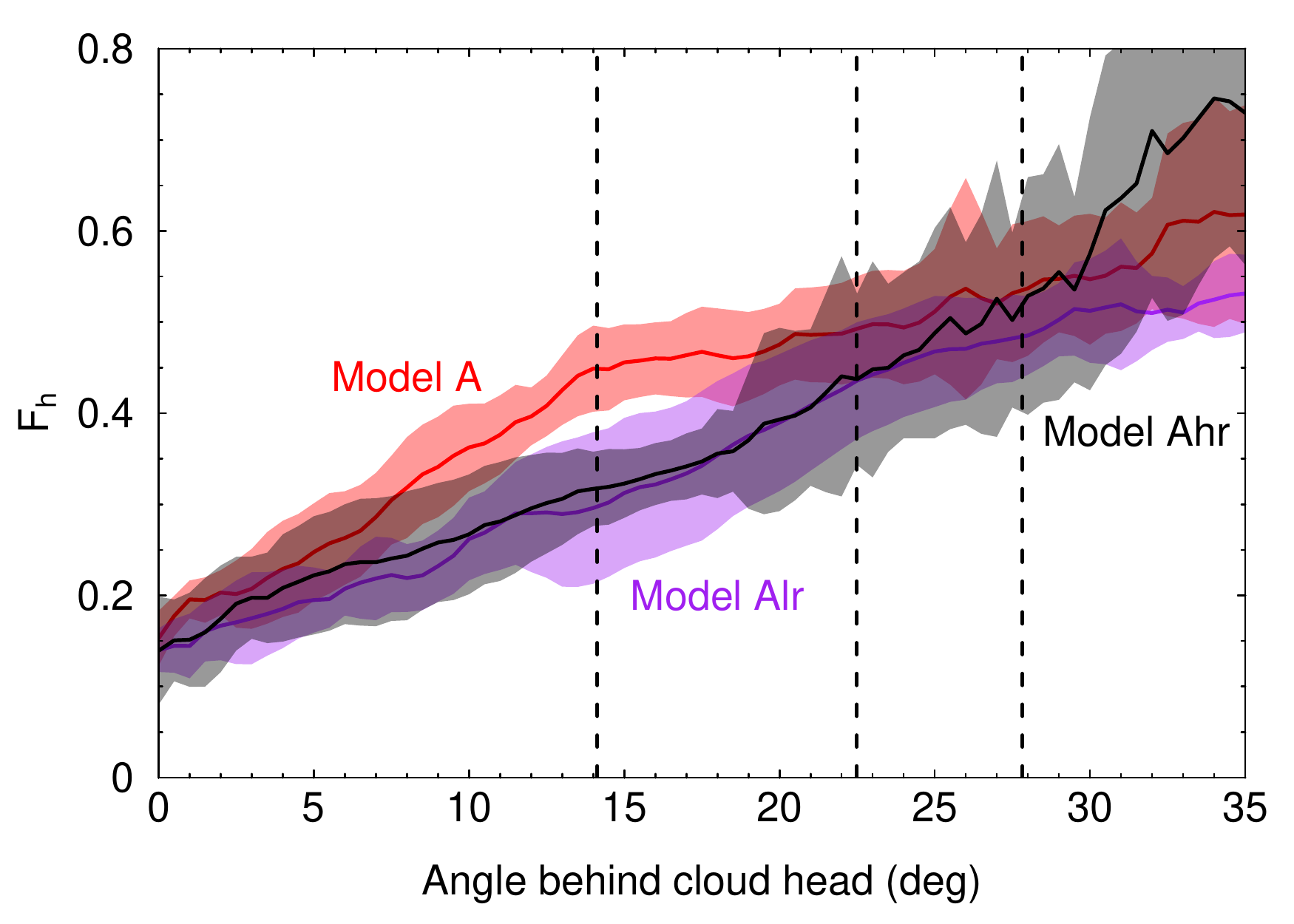}
\caption{Similar to Figure~\ref{fig:AverageProfile}, but comparing Models~Alr, A, and Ahr, which
  have maximum resolutions of 20, 10, and 5~pc, respectively (see Table~\ref{tab:ModelParameters}).
  \label{fig:CompareResolution}}
\end{figure}

\begin{figure*}
\centering
\plottwo{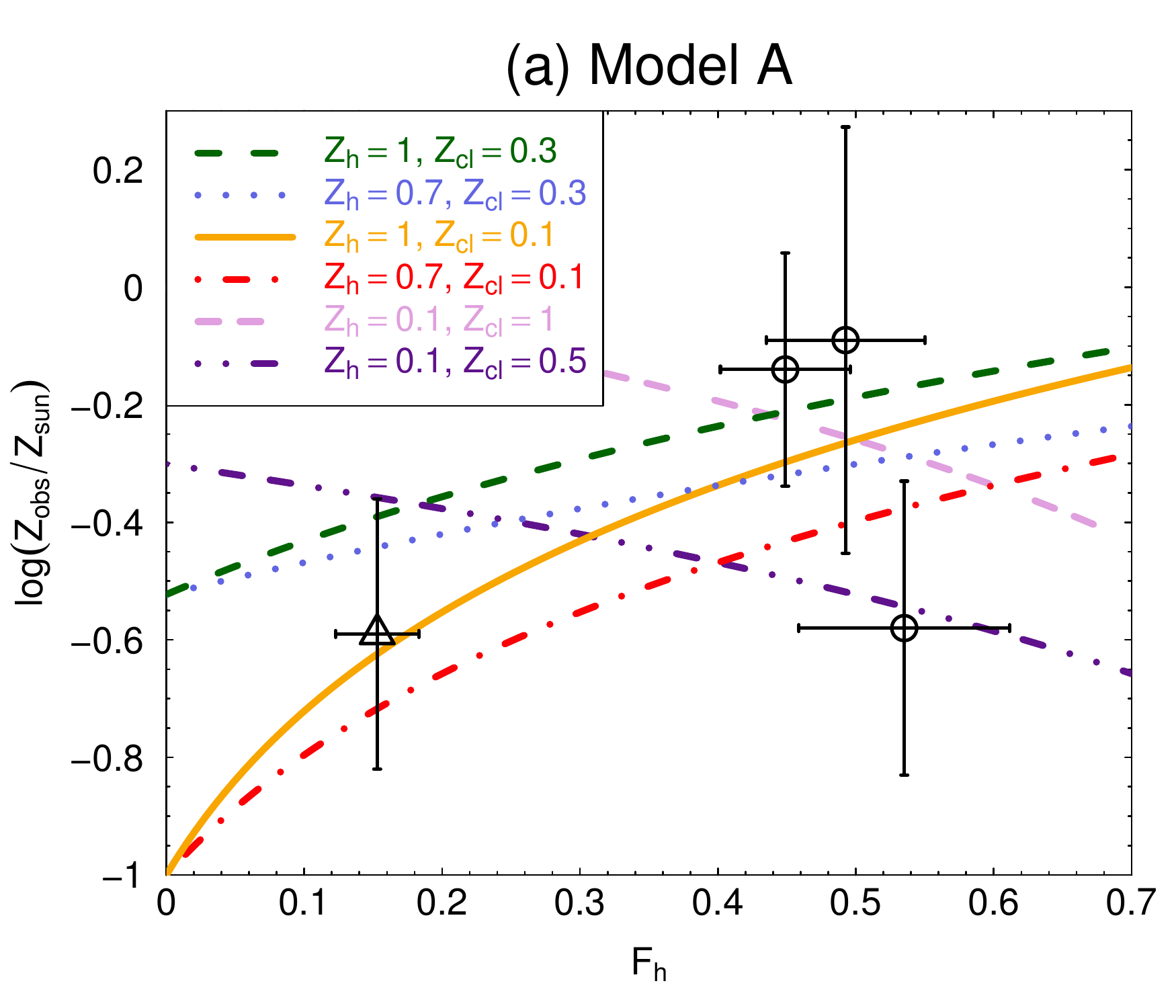}{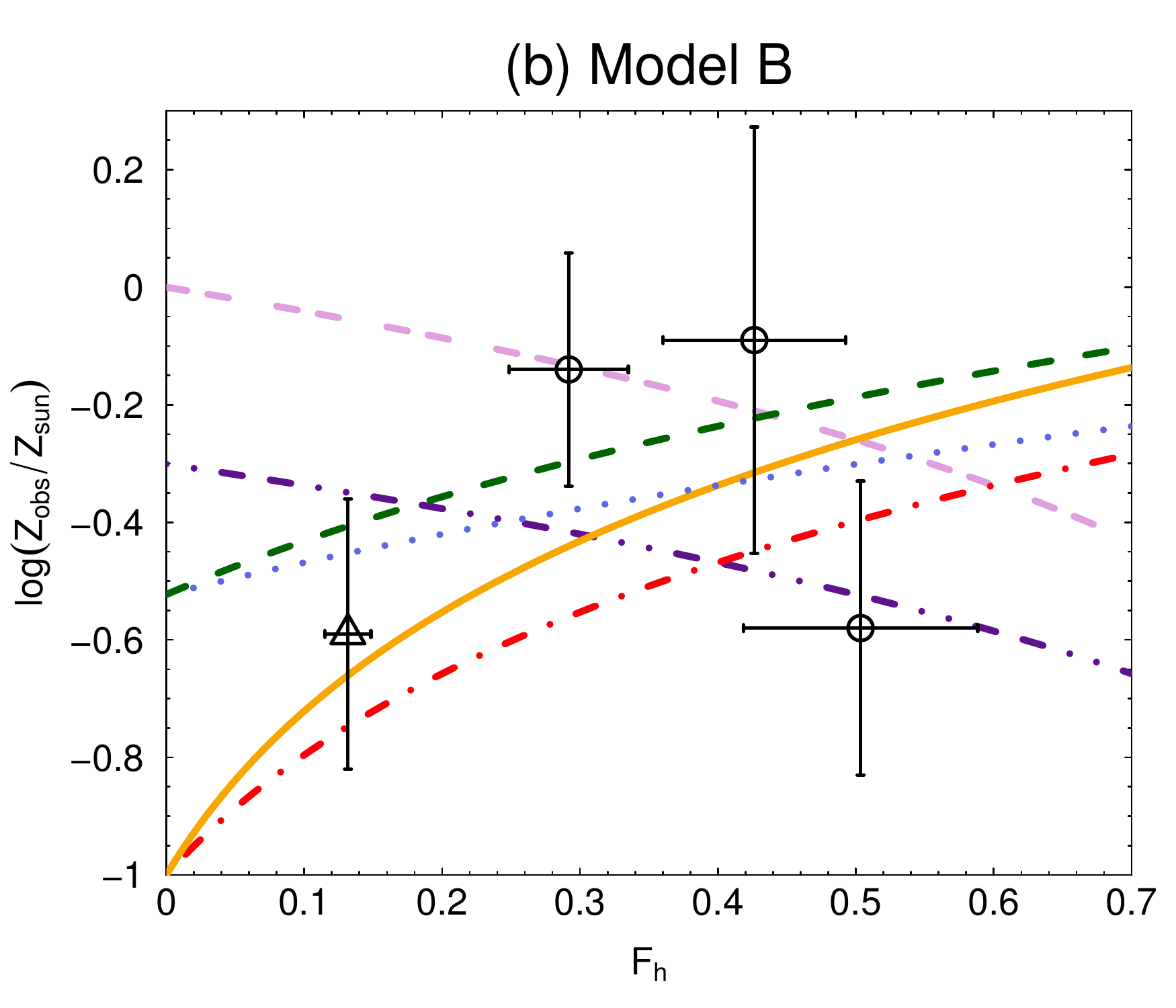}
\plottwo{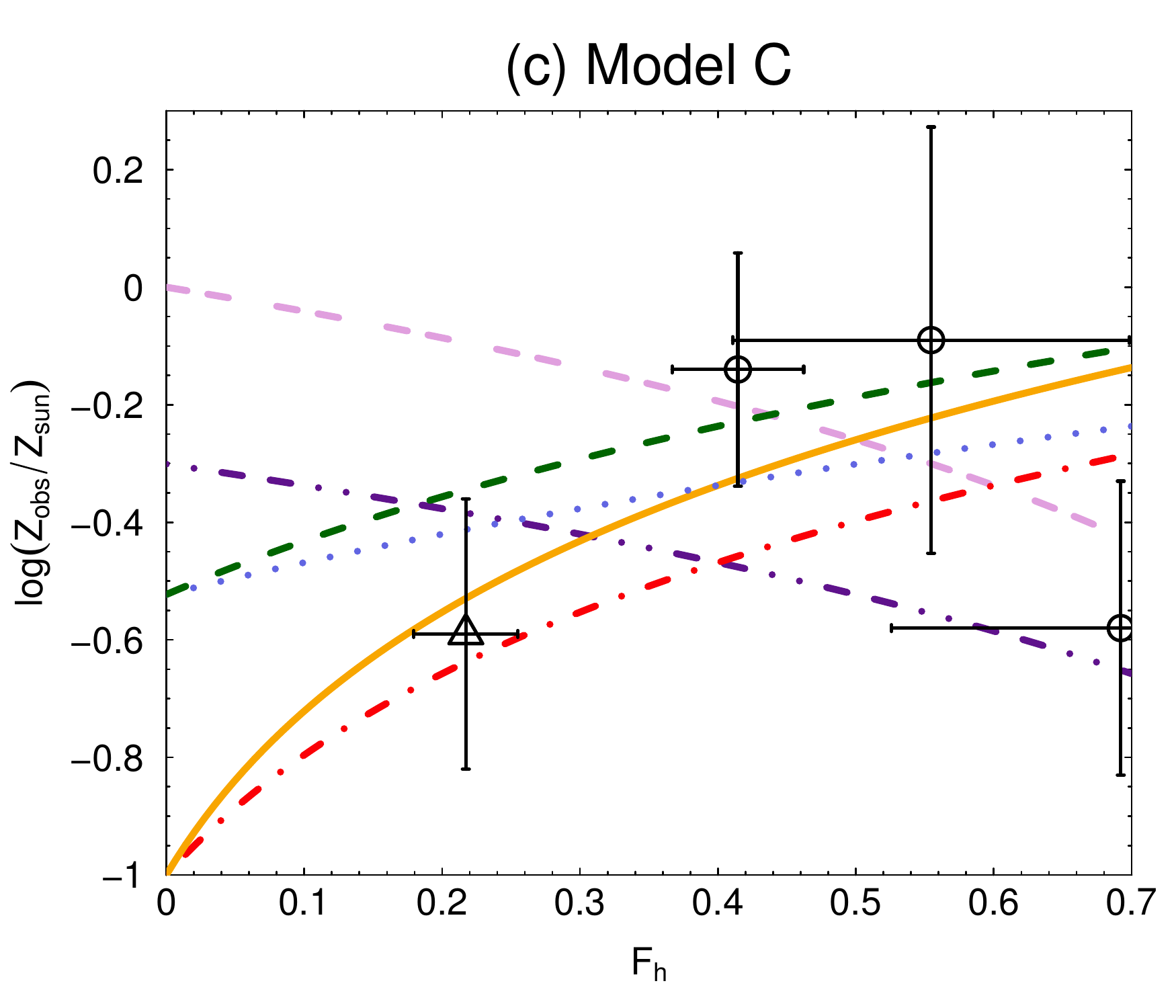}{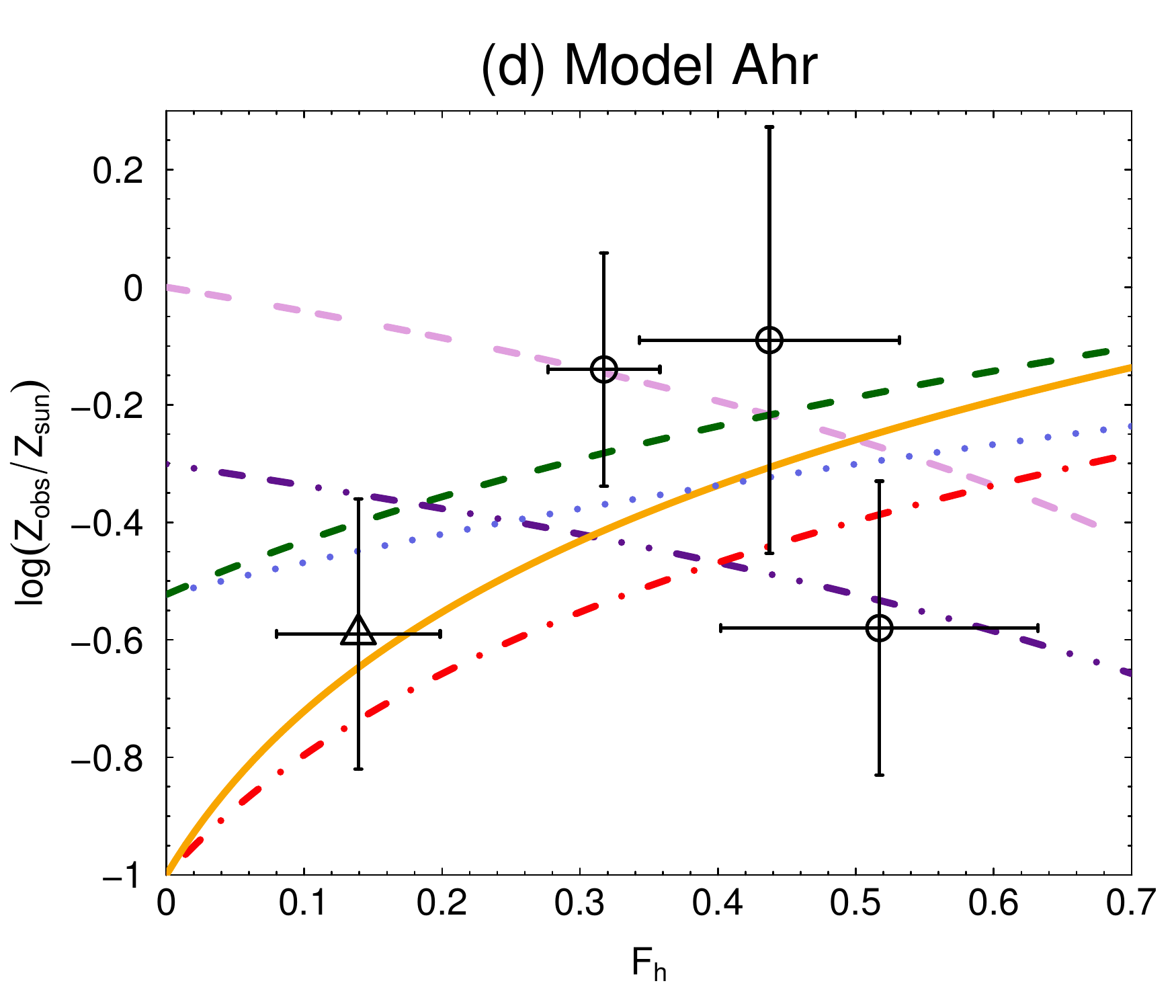}
\caption{Observed SC metallicities (triangle: H09; circles: F16) plotted against the values of
  \Fhalo\ that were derived from Models A--C and Ahr ((a)--(d), respectively) for the observed
  angular separations between the observed directions and the cloud's head. The error bars indicate
  $1\sigma$ uncertainties. The various curves show the relationships between \Fhalo\ and the
  metallicity of high-velocity material that would be expected to be observed for different values
  of \Zhalo\ and \Zcl\ (relative to solar), calculated using Equation~(\ref{eq:Zobs}).
  \label{fig:MetalvMixing}}
\end{figure*}

Figure~\ref{fig:CompareResolution} compares the averaged profiles calculated from versions of our
reference model with different maximum resolutions. In general, the low- and high-resolution
versions of the model (Alr and Ahr, respectively) yield similar results to each other. The
standard-resolution version also yields similar results, except for $\sim$7\degr--19\degr\ behind
the cloud head.  However, when we compare our model predictions with observations in the following
section, we find that this discrepancy does not affect our conclusions. We therefore conclude that a
maximum resolution of 10~pc (our standard resolution) is adequate for our purposes.

\section{COMPARISON WITH OBSERVATIONS}
\label{sec:Comparison}

The SC metallicity measurements from H09 and F16 are summarized in Table~\ref{tab:Obs}.  In
addition, we used the values shown in Figure~\ref{fig:AverageProfile} to estimate \Fhalo\ (the
fraction of the high-velocity material that originated in the halo) for synthetic sight lines that
cross through the cloud as far behind the SC head and at the same angles as each of the observed
sight lines do.  These values are shown in the final three columns of Table~\ref{tab:Obs}, for each
of our three models. The tabulated uncertainties on \Fhalo\ indicate the one standard deviation
variation in the model results.

Figure~\ref{fig:MetalvMixing} plots the metallicites observed by H09 and F16 versus the values of
\Fhalo\ we calculated from our models for equivalent lines of sight through the model SC.
Overplotted are the metallicities of high-velocity material that would be observed (\Zobs) as a
function of \Fhalo, given various possible values of intrinsic halo and cloud metallicities
(\Zhalo\ and \Zcl, respectively) and their relationship
\begin{equation}
  \Zobs = \Fhalo \Zhalo + (1 - \Fhalo) \Zcl.
  \label{eq:Zobs}
\end{equation}

In general, the predictions that assume a metal-rich halo ($\Zhalo\ga0.7$) and a metal-poor initial
cloud ($\Zcl\la0.3$) agree with the metallicity measurement for the head and two of the measurements
for the tail. However, the metallicity measurement from the sight line furthest down the tail
(RX~J2139.7+0246, the rightmost data point in Figure~\ref{fig:MetalvMixing}) is overpredicted.  It
is possible that the metallicity along the RX~J2139.7+0246 sight line is an outlier, as the
high-velocity \ion{H}{1} column density on this sight line is 4--5 times those on F16's other two
sight lines, despite its being the furthest from the SC head.  It is also possible that this sight
line passes through a relatively pristine (and therefore low-metallicity, in the scenario considered
here) blob of cloud material that was stripped from the head of the cloud without undergoing
significant mixing. Such pristine blobs are sometimes seen in the tails of model HVCs
(J.~A. Gritton, 2016, private communication).  Alternatively, the nonmonotonic trend in observed
metallicities may indicate spatial variation in the halo or cloud metallicities. Excluding the
RX~J2139.7+0246 sight line, the remaining three data points imply $0.1\la\Zcl\la0.3$ and
$0.7\la\Zhalo\la1.0$.  If the halo metallicity is near solar, then $\Zcl\la0.3$ should be preferred
over $\Zcl\sim0.5$ (the average of the observed metallicities on the four sight lines), as the
latter would result in an even higher metallicity along the sight line furthest along the tail.

The predictions in the scenario that assumes a metal-poor halo ($\Zhalo\sim0.1$) and a metal-rich
initial cloud ($\Zcl\ga0.5$) also agree with some but not all of the data points. In particular, if
the cloud metallicity is large enough to match the measured metallicity on the RX~J2043.1+0324 and
PG~2112+059 sight lines (the middle two data points in Figure~\ref{fig:MetalvMixing}), then the
metallicity toward the cloud head (the leftmost data point) is overpredicted. However, a metal-poor
halo is inconsistent with independent measurements ($\Zhalo\ga0.6$; \citealt{miller16}).

Despite the differences between the various models apparent in Figures~\ref{fig:AverageProfile} and
\ref{fig:CompareResolution}, the above conclusions are the same for all models (hence our comment in
Section~\ref{sec:Mixing} that our maximum model resolution of 10~pc appears to be adequate).

\section{DISCUSSION AND CONCLUSIONS}
\label{sec:Discussion}

Our three-dimensional hydrodynamical simulations of an SC-like cloud passing through halo gas reveal
that the tail of the cloud entrains more halo gas than does the head of the cloud.  This naturally
results in an increase in metallicity from cloud head to tail if the halo has greater metallicity
than the cloud (and the reverse if the halo has lesser metallicity than the cloud).  The predictions
from both scenarios agree with the observed metallicites on some but not all of the observed sight
lines; however, the former scenario is preferred as it is in better accord with the findings of
\citet{miller16} that $\Zhalo\ga0.6$.

We explored models with different cooling curves, to represent the uncertainty in the halo and cloud
metallicities (our model does not calculate the metallicity dependence of the cooling
self-consistently). We also explored models with different ambient densities, representing the
change in the ambient density as the SC moves through the halo. While the general conclusions
derived from the various models are the same, the differences between the models contribute to the
uncertainty in the degree of mixing between the halo and the cloud (as does the time variation
within each model). As a result, the uncertainties on the inferred halo and initial cloud
metallicities are large.

Our models do not include magnetic fields. Magnetohydrodynamic (MHD) simulations suggest that
magnetic fields suppress mixing between cloud and halo material \citep{mccourt15}.  This suppression
of mixing would tend to shift the model predictions in Figure~\ref{fig:MetalvMixing} to the left,
meaning that the observed metallicity, \Zobs, would vary more rapidly with respect to the degree of
mixing, \Fhalo. Since, from Equation~(\ref{eq:Zobs}), $d\Zobs/d\Fhalo=\Zhalo-\Zcl$, this in turn
means that the difference between the halo and initial cloud metallicities inferred from MHD
simulations would be larger than from hydrodynamical simulations such as those presented here.

Additional observations could aid in determining the metallicity gradient along the SC.  Given the
current state of knowledge, we conclude that due to mixing with the halo, the SC's true original
metallicity may be less than is implied by H09's and F16's metallicity measurements.  Furthermore
its uncertainty is likely greater than is implied by the uncertainties on H09's and F16's
metallicity measurements.  Consequently, the ultimate origin of the SC remains uncertain.

\acknowledgements
We wish to thank Jason Galyardt for assistance with programming and F.\ Jay Lockman, Andrew
Fox, Alex Hill, J.\ Chris Howk, and Nicolas Lehner for interesting conversations about the Smith
Cloud and other HVCs. We also thank the anonymous referee whose comments helped improve
this paper.
The software used in this work was developed in part by the DOE NNSA ASC- and DOE Office of Science
ASCR-supported Flash Center for Computational Science at the University of Chicago.
The simulations were performed at the Georgia Advanced Computing Resource Center (GACRC) of
the University of Georgia. We thank Shan-Ho Tsai for her invaluable technical support.
We acknowledge use of the R and yt software packages \citep{R,turk11}.
This research was funded by NASA grant NNX13AJ80G, awarded through the Astrophysics Theory Program.

\end{document}